# Non-Isolated Single-Switch Zeta Based High-Step up DC-DC Converter with Coupled Inductor

Armin Abadifard
*Faculty of Electrical & Computer Engineering*
*University of Tabriz*
Tabriz, Iran
a.abadifard93@ms.tabrizu.ac.ir

Pedram Ghavidel
*Faculty of Electrical & Computer Engineering*
*University of Tabriz*
Tabriz, Iran
p.ghavide@ms.tabrizu.ac.ir

Seyed Hossein Hosseini
*Electrical and Computer Engineering Department*
*University of Texas at Dallas*
Richardson, USA
seyedhossein.hosseini@utdallas.edu

Masoud Farhadi
*Electrical and Computer Engineering Department*
*University of Texas at Dallas*
Richardson, USA
Masoud.Farhadi@utdallas.edu

*Abstract*—In this paper, a non-isolated high step-up DC-DC converter has been proposed for renewable energy applications. The proposed structure converter has been derived from the fundamental Zeta converter, in both of which only a single switch is employed. The voltage gain ratio has considerably enhanced in this converter with the absence of using switched capacity. To magnify voltage gain of the converter, a coupled-inductor has adopted. Increase and decrease of gain by changing the ratio of coupled-inductor assist the duty cycle. The number of components is low in this structure. The operating principle and evaluation of the proposed converter, considering designing approaches for elements, are discussed in detail. To verify the feasibility of the proposed converter, simulation results have been provided and evaluated.

## I. Introduction

It goes without saying that in such a sophisticated world where the realms of knowledge are expanding dramatically, providing electric energy for the consumers is one of the main concerns for electrical engineers. That is the reason why today, we rely heavily on renewable energies. Such clean energies are highly prominent in the industry due to the fact that they are: environmentally friendly energies, accessible, quite high in amount, and reliable [1]-[2]. Distributed generation systems such as photovoltaic systems, wind turbines, fuel cells are considered as clean energy sources, in all of which micro sources have been used [3]-[5]. Generally, the generated DC voltages by micro sources are low and need to be enhanced before being converted to the AC. In this regard, high step-up DC-DC converters are playing a crucial role in being engaged as a strong DC link to provide a desirable voltage level for micro sources, and finally, desirable AC voltage [6]-[7]. The main reasons for having a low voltage gain ratio in power converters are:

a) The operation of power switches and diodes
b) Equivalent resistance circuit (Resistance, inductors, capacitors .i.e.)

To achieve a higher voltage value, embedding voltage cells in series in photovoltaic or using step-up DC-DC converters such as boost, fly-back are practical approaches. In boost converters, an increase in the value of the duty cycle yields a high output voltage. However, increasing the duty cycle would bring the following disadvantages:

a) It would put the switching performance in jeopardy.
b) Increases the power losses made by power switches.
c) Electromagnetic interference (EMI) issues.
d) For high duty cycles in boost converters, the dynamic response in small-signal analysis would face severe limitations [8]-[9].

Arranging PV cells in series to achieve a desirable and higher DC voltage is one of the significant concerns under the effect of shadow to reach maximum power point track in PV systems [2], [5]. An increase in DC voltage in isolated converters is possible by turning the ratio of transformer. However, the effect of leakage flux of the transformer would damage the power switches [10]. Fly-back converters can enhance the DC voltage. Nevertheless, they suffer from high power loss and high voltage stress imposed on power switches. Ostensibly, Snubber. and active clamp circuits can mitigate such problems; but they increase the costs of implementation, which is a negative factor in the power electronics area [11], [19]-[23]. In this regard, various topologies of non-isolated high step-up converters have offered to guarantee the high DC voltage transfer without the need to increase much of the duty cycle. A suitable converter must have the following advantages:

a) High efficiency
b) Offering a high voltage gain ratio.

Recent investigations using switched-capacitor configuration, voltage lift technique, coupled-inductor mainly focused on bringing the advantage of high voltage gain ratio in their structure [12]. Adopting a switched-capacitor would significantly reduce the voltage stress across the semiconductors (mostly power switches) in a converter [13]-[18]. Nevertheless, due to the flow of the transient current in power switches, the total power loss of the converter will increase, which affects the maximum efficiency of such power converters. Therefore, engaging the switched capacitor architectures to a converter in the presence of coupled-inductor would be a quite significant contribution to reach a higher DC voltage gain at the output.

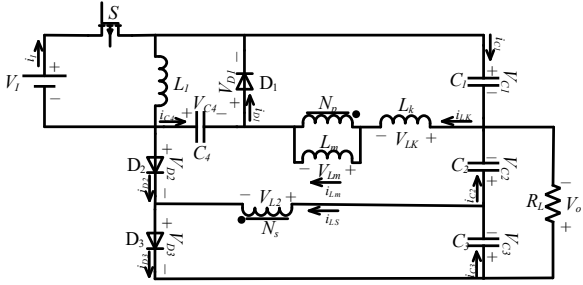

Fig. 1. The topology of the proposed converter

Taking all the mentioned concepts into the account, this study proposes a single-switch high-step up DC–DC converter. This converter has derived from the topology of the fundamental Zeta converter. The proposed structure uses a coupled-inductor, hence charges the capacitor with the use of energy stored in the coupled-inductor. Also, the coupled-inductor considerably improves the voltage gain. Only a single power switch has been used to attain a simple structure with a simple operating principle. This structure has a low element in comparison with similar converters.

The following of this paper have arranged as follows. Section II discusses the operation principles of the proposed converter. Section III is devoted to the steady-state calculations mathematical. Section IV provides a comprehensive parameter design for proposed converter. In Section V, the simulation results discussed in detail. Finally, Section VI concludes the findings of this research.

## II. OPERATION PRINCIPLE OF THE PROPOSED CONVERTER

The topology of the proposed converter is demonstrated in Fig. 1. In this circuit, a DC power source $V_{in}$ in the input and a resistive load $R_L$ at the output is engaged. The proposed converter has a power switch S, three diodes, four capacitors, and a single inductor. The capacitor $C_1$ and the diode $D_1$ are serving as clamp circuit in this converter. The capacitors $C_2$ and $C_3$ not only are used as output capacitors but also are employed as the capacitors of the extended voltage multiplier cells (VMCs). The coupled-inductor is modeled by the magnetizing current of $L_m$ and the leakage flux of $L_K$, where k is the coupling coefficient of the coupled-inductor. The ratio of the primary and secondary winding of the coupled-inductor is defined as $N_P/N_S$. To simplify the steady-state calculations, ideal behavior of all components are assumed by considering the following critical assumptions:

1) All elements are treated as an ideal without any parasitic effects.
2) All passive elements, such as capacitors and inductors are considered as ripple-free. In other words, the voltage across all capacitors and the current through all inductors, are constant during a switching interval.

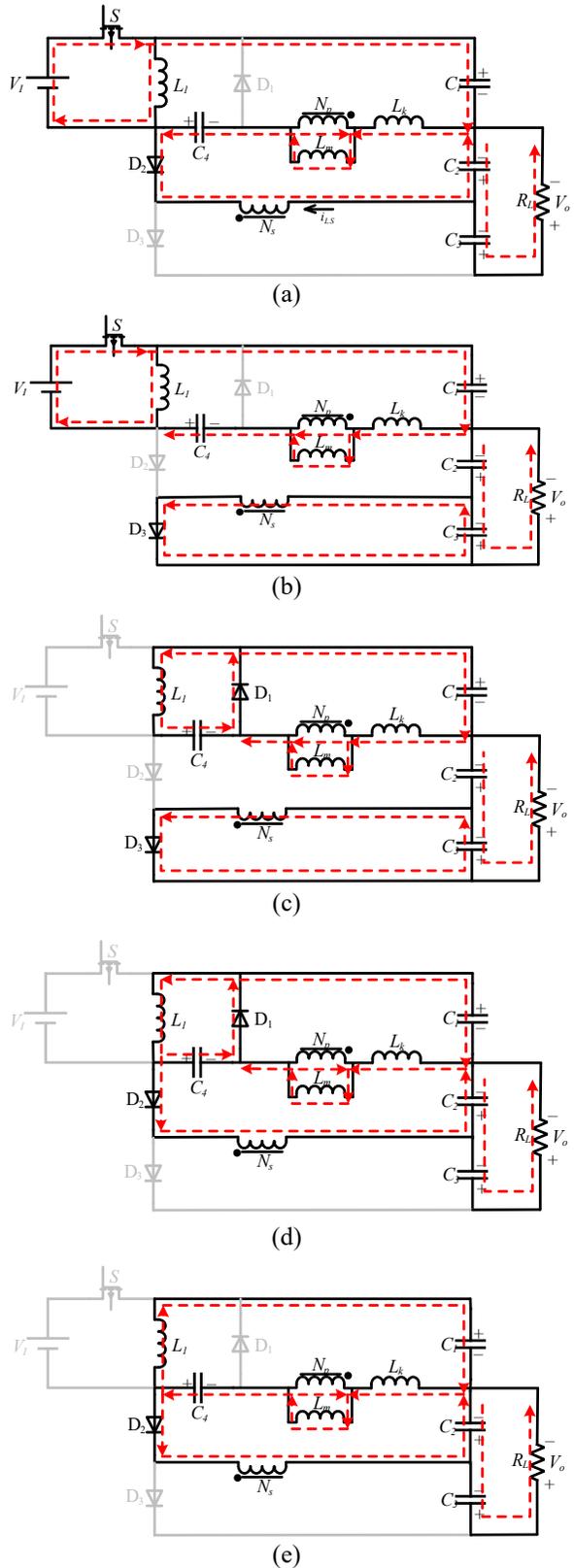

Fig. 2. The current flow directions in the proposed converter under CCM operation, (a) Mode 1, (b) Mode 2, (c) Mode 3, (d) Mode 4, (e) Mode 5.

Considering the abovementioned conditions, the proposed converter has five distinct modes of operation under continuous conduction mode (CCM). The equivalent circuit of all modes are displayed in Fig. 2 (a)-(e). Based on Fig. 2, the typical time-domain waveforms of the proposed converter are shown in Fig. 3, all under CCM operation. The five intervals can be defined as follows:

Mode 1[$t_0$-$t_1$]: According to Fig. 2 (a), the power switch is conducting, and $D_2$ is turned on, while $D_1$ and $D_3$ are reversed biased. Accordingly, both the leakage flux current of $L_k$ and the current through the secondary side of the coupled-inductor are increased, whereas $L_m$ is decreased linearly. This situation continues until $L_k$ becomes equal to $L_m$, and the current through the secondary side of the coupled-inductor meets zero. Moreover, $L_1$ is magnetized by the input source.

Mode 2[$t_1$-$t_2$]: As depicted in Fig. 2 (b), the power switch is still on, and $D_1$ is off. When $i_{Lk}$ reaches $i_{Lm}$, $D_2$ becomes off, and the second mode begins. At this moment, $i_{Lk}$ is still increasing, and the difference of $i_{Lm}$ and $i_{Lk}$ is transferred to the secondary side of coupled inductor, so, $D_3$ become on. The current through $L_1$, is increased because the $S$ is on. This interval continues until the power switch becomes off.

Mode 3[$t_2$-$t_3$]: As shown in Fig. 2 (c), at first, $S$ is off, and $i_{L1}$ turns $D_1$ on. The diode $D_2$ is in off mode, while $D_3$ is on. Meanwhile, all inductors start to demagnetize their energy to charge $C_1$ and $C_2$. This process is fast because the leakage energy of coupled-inductor are relatively low.
This energy decreases with a high slope. As $i_{Lk}$ stands equal to $i_{Lm}$, the current through the secondary side of the coupled-inductor meets zero, and $D_3$ becomes reverse biased. So, this interval is finished.

Mode 4[$t_3$-$t_4$]: As demonstrated in Fig. 2 (d), $D_1$ and $D_2$ are on, whereas $S$ and $D_3$ are off. The current of $L_m$ is decreased with the different slope compared to Mode 3. This process continues until $i_{Lm}$ becomes $-i_{L1}$. Meanwhile, the energy for the output load is provided by $C_2$ and $C_2$. Moreover, $i_{Lm}$ is decreased in this interval.

Mode 5[$t_4$-$t_5$]: According to Fig. 2 (e), $D_1$ and $S$ are reversed biased. The currents $i_{Lm}$ and $i_{L1}$ are equal. Meanwhile, $C_2$ is charged by the secondary side of the coupled-inductor to fulfill $R_L$. This inverval continues until the $S$ is on.

### III. STEADY-STATE ANALYSIS

The proposed converter is analyzed under the CCM operation. To simplify the calculation of leakage inductance, both Modes 1 and 3 are ignored. Thence, in Mode 2, we have:

$$V_{L1} = V_I \quad (1)$$
$$V_{Lm} = V_I + V_{C4} - V_{C1} \quad (2)$$
$$V_{C3} = V_{Ls} = nV_{Lm} \quad (3)$$
$$V_{C3} = n(V_I + V_{C4} - V_{C1}) \quad (4)$$

In equation (3), $n$ represents the ratio of the coupled-inductor. In the Modes 4 and 5, the following equations can be derived:

$$V_{L1} = -V_{C4} \quad (5)$$
$$V_{Lm} = -V_{C1} \quad (6)$$
$$nV_{C1} + V_{C4} + V_{C1} - V_{C2} = 0 \Rightarrow$$
$$(n+1)V_{C1} + V_{C4} = V_{C2} \quad (7)$$

From (5)-(7), the full expression for a complete switching period can be achieved as follows:

$$V_{L1} = \begin{cases} V_I & D \\ -V_{C4} & 1-D \end{cases} \quad (8)$$

$$V_{Lm} = \begin{cases} V_I + V_{C4} - V_{C1} & D \\ -V_{C1} & 1-D \end{cases} \quad (9)$$

From (8) and (9), one can conclude that

$$V_{C4} = \frac{D}{1-D}V_I \quad (10)$$

$$V_{C1} = D(V_I + V_{C4}) \quad (11)$$

By using (10)-(11), we have:

$$V_{C1} = \frac{D}{1-D}V_I \quad (12)$$

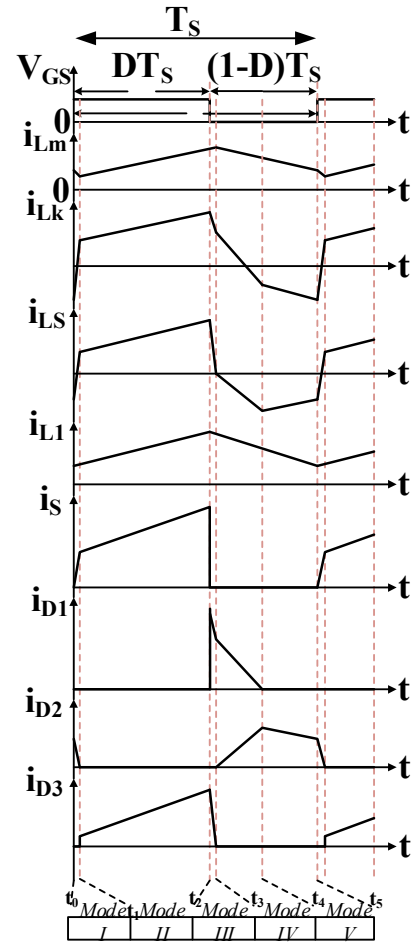

Fig. 3. Typical time-domain waveforms of the proposed converter under CCM operation.

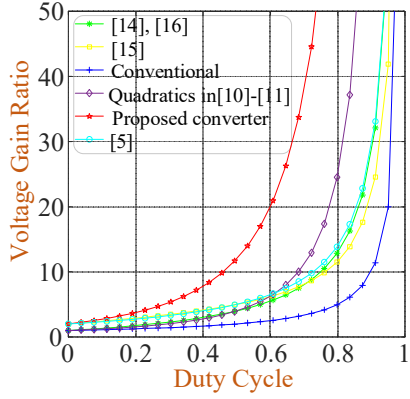

Fig. 4. Voltage gain ratio comparison.

The voltage across $C_2$ can be achieved by using (7), (10), and (12) as follows.

$$V_{C2} = \frac{(n+2)D}{1-D}V_I \qquad (13)$$

The voltage across $C_3$ can be earned by using (4), (11), and (12) as follows.

$$V_{C3} = nV_I \qquad (14)$$

The output voltage can be defined by

$$V_o = V_{C2} + V_{C3} \qquad (15)$$

Finally, by substituting (13)-(14) to (15), the voltage gain ratio of the proposed converter can be written as

$$M = \frac{n+2D}{1-D} \qquad (16)$$

A comparison has made between the proposed converter and relevant structures, where are given in Table I. Based on this table, the voltage gain curves versus duty cycle are plotted and shown for the proposed converter and that of the conventional boost converter and given in [5], [14], [15], [16] and quadratic structures in [10]-[11]. As shown in Fig 4, the proposed converter has a higher voltage gain ratio (n=2) as the duty cycle increases. Moreover, the proposed converter has a relatively lower component count in comparison to the abovementioned competitors. It is of interest to note that the proposed converter uses a single power switch, which brings the advantage of easy controlling ability.

By applying KVL on the equivalent circuit of the proposed converter in Figs. 2, the voltage stress across diodes and the main power switch are achieved by:

$$V_{D1} = V_S = -\frac{V_I}{1-D} \qquad (17)$$

$$V_{D2} = -\frac{(1+n)V_I}{1-D} \qquad (18)$$

$$V_{D3} = -\frac{nV_I}{1-D}. \qquad (19)$$

According to Figs. 2, the average input current is:

$$I_{in(ON)} = DI_{L1} + DI_{Lm} + I_{D2} \qquad (20)$$

$$I_{in(off)} = 0 \qquad (21)$$

Using (20)-(21), the average of magnetizing current can be given by

$$I_{Lm} = \frac{I_O(2D+n-1)}{D} \qquad (22)$$

Now using integral equation, the current of $L_1$ and $L_m$ are

$$i_{L1}(t) = i_{L1}(t_0) + \frac{1}{L_1}\int_0^t v_{L1}(\tau)d\tau \qquad (23)$$

$$i_{Lm}(t) = i_{Lm}(t_0) + \frac{1}{L_m}\int_0^t v_{Lm}(\tau)d\tau \qquad (24)$$

Assuming $t_0=0$, and $t=DT$, the current ripple of $L_m$ and $L_1$ is found by

$$\Delta i_{L1} = \frac{DV_{in}}{L_1 f_s} \qquad (25)$$

$$\Delta i_{Lm} = \frac{DV_{in}}{L_m f_s} \qquad (26)$$

Based on Fig. 3 and applying ampere-second balance on all capacitors, one can conclude that the average current through diodes equals the average current of $R_L$. In this regard, the peak current through each diode and the main power switch can be earned by

$$i_{D3(peak)} = \frac{2I_O}{D} \qquad (27)$$

$$i_{D2(peak)} = \frac{2I_O n}{D(n+1)} \qquad (28)$$

$$i_{D1(peak)} = i_{S(peak)} = \frac{2nI_O}{D} + \frac{DV_{in}}{f_s L_m} + \frac{DV_{in}}{L_1 f_s}$$
$$= I_O\left(\frac{2n}{D} + \frac{DR_L}{f_s M}(L_1 \quad L_m)\right) \qquad (29)$$

The length of the third and fourth intervals can be calculated by the following equation.

$$d_3 + d_4 = \frac{2}{\frac{2n}{D} + \frac{DR_L(1-D)}{f_S(L_m \quad L_1)(n+D)}} \qquad (30)$$

IV. DESIGN OF PASSIVE COMPONENTS

According to [17], the following assumptions must be made for the converter to operate in CCM.

$$I_{Lm} \geq \frac{1}{2}\Delta i_{Lm} \qquad (31)$$

Using (19), (21), and (26), the minimum value for the current through $L_m$ is

$$L_m \geq \frac{D^2 V_{in}}{2(2D+n-1)I_O f_S} \qquad (32)$$

Since the output capacitors provide the energy for $R_L$, and also they charge the primary winding, they play an important role in voltage ripples of the system. In this regard, the worst case for designing the capacitors must be considered as given in follows.

TABLE I
GENERAL INFORMATION OF THE PROPOSED CONVERTER AND
COMPETITIVE BOOST CONVERTER

| Competitors | Components Count | Voltage Gain Ratio |
|---|---|---|
| Convectional boost converter | 1 Switch<br>1 Diode<br>1 Capacitor<br>1 inductor | $\dfrac{1}{1-D}$ |
| [5] | 1 Switch<br>4 Diode<br>5 Capacitors<br>1 inductor | $\dfrac{2+D}{1-D}$ |
| [14] | 2 Switches<br>4 Diodes<br>3 Capacitors<br>3 inductors+CP | $\dfrac{1+2D}{1-D}$ |
| [15] | 2 Switches<br>4 Diodes<br>3 Capacitors<br>3 inductors+CP | $\dfrac{2}{1-D}+nD$ |
| [16] | 1 Switch<br>2 Diodes<br>4 Capacitors<br>3 inductors | $\dfrac{1+nD}{1-D}$ |
| Quadratics in [10]-[11] | 2 Switch<br>2 Diodes<br>2 Capacitor<br>2 inductors | $\dfrac{1}{(1-D)^2}$ |
| Proposed converter | 1 Switches<br>3 Diode<br>4 Capacitors<br>1 Inductors+CP | $\dfrac{n+2D}{1-D}$ |

$$C_{1\min} = C_{4\min} = \frac{1}{V_{PPC}f_S}\left(\frac{D^2}{L_m f_S} + \frac{2n(n+2D)}{R_L(1-D)}\right) \quad (33)$$

$$C_{2\min} = \frac{2n(n+2D)}{V_{PPC}f_S R_L D(1+n)} \quad (34)$$

$$C_{3\min} = \frac{2(n+2D)}{V_{PPC}f_S R_L (1-D)} \quad (35)$$

## V. MATLAB SIMULATION RESULTS

In this section, the proposed converter has been examined and evaluated by a simulation made in MATLAB Simulink software. The values of passive components are $C_1=C_4=47\mu F$, $C_2=C_3=3.3\mu F$, $L_1=47\mu H$. Also, $L_K$: 1 µH, $L_m$: 300 µH. These values have been considered based on the current ripples of inductors and voltage ripples across capacitors that all have determined in the previous section. The switching frequency of the proposed converter is considered as 40kHz. The waveforms are detected and demonstrated in Figs. 5 (a)-(h). In this condition, the proposed converter receives energy from a 30V input DC power source. By choosing D=60% and n=2, which yield the voltage gain of 8, which is in coincided with the displayed waveform and also with the equation (9). The critical waveforms such as: $V_o$, $I_L$, $V_S$, $I_S$, $V_{D1}$, $I_{D1}$, $V_{D2}$, $I_{D2}$, $V_{D3}$, $I_{D3}$, $I_{L1}$, $I_{Lm}$ $I_{Lk}$, and $I_{Lm}$ in CCM operation are demonstrated.

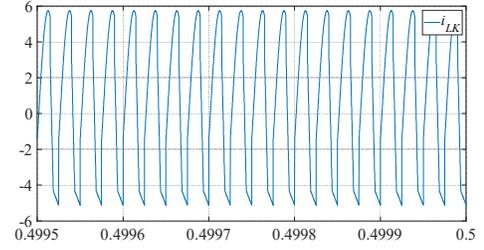
(a)

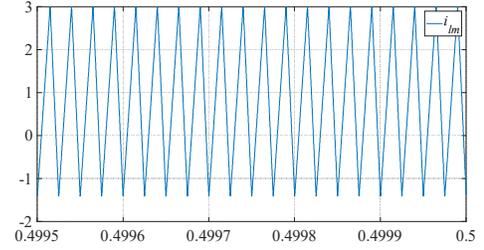
(b)

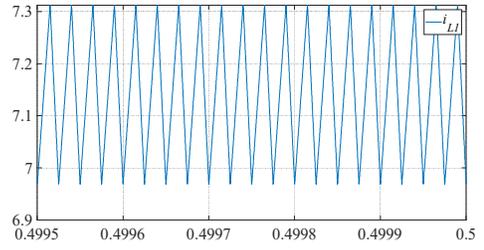
(c)

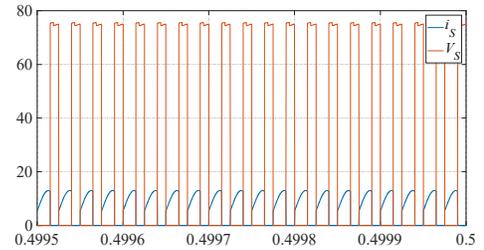
(d)

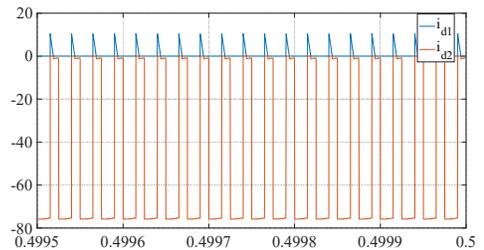
(e)

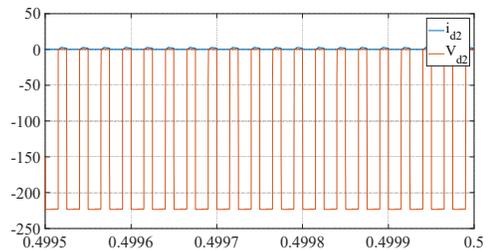
(f)

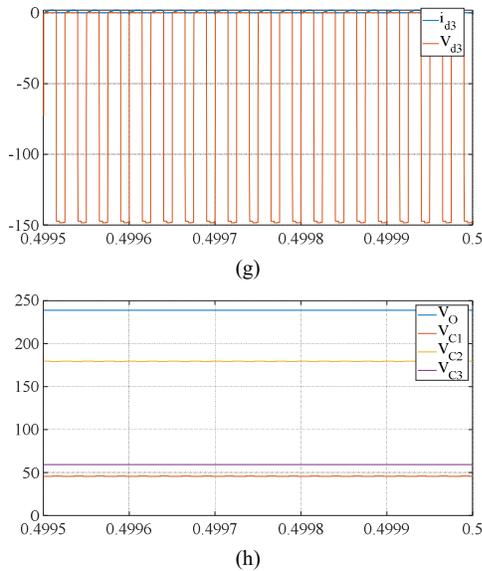

Fig. 5. Simulation results of the proposed converter under CCM operation.

## VI. Conclusion

In this paper, a novel topology for a non-isolated single-switch zeta based high-step up DC-DC converter with a coupled-inductor is presented. Mathematical approaches have been made to derive the steady-state equations of the proposed converter under CCM operation. A comprehensive comparison has been made between the proposed converter and main competitive structures. As a result, the proposed converter has a high voltage gain ratio, lower component counts among its similar structures using only a single switch. The values considered for the proposed converter have been designed. Simulation results have demonstrated the feasibility of the proposed converter. In conclusion, all the detected simulations and experimental waveforms are in agreement with the expected time-domain waveforms.